\documentclass[a4paper,fleqn]{cas-dc}

\usepackage[numbers]{natbib}

\def\tsc#1{\csdef{#1}{\textsc{\lowercase{#1}}\xspace}}
\tsc{WGM}
\tsc{QE}
\tsc{EP}
\tsc{PMS}
\tsc{BEC}
\tsc{DE}

\AtBeginDocument{\providecommand\subsecref[1]{\ref{subsec:#1}}}

\begin{document}
\let\WriteBookmarks\relax
\def\floatpagepagefraction{1}
\def\textpagefraction{.001}
\shorttitle{Unveiling the role of plasticity rules in RC}
\shortauthors{G. B. Morales et~al.}

\title [mode = title]{Unveiling the role of plasticity rules in reservoir computing}                      

\author[1,2]{Guillermo B. Morales}[orcid=0000-0002-6769-7979]
\ead{guillermobm@onsager.ugr.es}


\address[1]{Instituto de F\'{i}sica Interdisciplinar y Sistemas Complejos (IFISC, UIB-CSIC), Campus Universitat de les Illes Balears E-07122, Palma de Mallorca, Spain}

\author[1]{Claudio R. Mirasso}[orcid=0000-0003-2980-7038]
\ead{claudio@ifisc.uib-csic.es}

\author[1]{Miguel C. Soriano}[orcid=0000-0002-6140-8451]
\cormark[1]
\ead{miguel@ifisc.uib-csic.es}



\address[2]{Instituto Carlos I de F\'{i}sica Te\'{o}rica y Computacional, Facultad de Ciencias, Campus de Fuentenueva 18071 Granada, Spain}

\cortext[cor1]{Corresponding author}

\begin{abstract}
Reservoir Computing (RC) is an appealing approach in Machine Learning that combines the high computational capabilities of Recurrent Neural Networks with a fast and easy training method. Likewise, successful implementation of neuro-inspired plasticity rules into RC artificial networks has boosted the performance of the original models. In this manuscript, we analyze the role that plasticity rules play on the changes that lead to a better performance of RC. To this end, we implement synaptic and non-synaptic plasticity rules in a paradigmatic example of RC model: the Echo State Network. Testing on nonlinear time series prediction tasks, we show evidence that improved performance in all plastic models are linked to a decrease of the pair-wise correlations in the reservoir, as well as a significant increase of individual neurons ability to separate similar inputs in their activity space. Here we provide new insights on this observed improvement through the study of different stages on the plastic learning. From the perspective of the reservoir dynamics, optimal performance is found to occur close to the so-called edge of instability.

 Our results also show that it is possible to combine different forms of plasticity (namely synaptic and non-synaptic rules) to further improve the performance on prediction tasks, obtaining better results than those achieved with single-plasticity models.
\end{abstract}

\begin{keywords}
Reservoir Computing \sep Plasticity \sep Hebbian
learning \sep Intrinsic plasticity \sep Nonlinear time series prediction.
\end{keywords}

\maketitle

\section{Introduction}

From the first bird-inspired ``flying machines'' of Leonardo da
Vinci to the latest advances in artificial photosynthesis, humankind
has constantly sought to mimic nature in order to solve complex problems.
It is therefore not surprising that the dawn of Machine Learning (ML)
and Artificial Neural Networks (ANN) was also characterized by the
idea of emulating the functionalities and characteristics of the human
brain. Within his book \textit{The Organization of Behavior}, Donald
Hebb proposed in 1949 a neurophysiological model of neuron interactions
that attempted to explain the way associative learning takes place
\citep{hebb_organization_1949}. Theorizing on the basis of synaptic
plasticity, Hebb suggested that the simultaneous activation of cells
would lead to the reinforcement of the involved synapses, a hypothesis
often summarized in the today's well-known statement: ``neurons that
fire together, wire together''. Thus, Hebbian theory was swiftly
taken by neurophysiologists and early brain modelers as the foundation
upon which to build the first working artificial neural network. In
1950, Nat Rochester at the IBM research lab embarked in the project
of modeling an artificial cell assembly following Hebb's rules \citep{milner_brief_2003}.
However, he would soon be discouraged by an obvious flaw in Hebb's
initial theory: as connection strength increases with the learning
process, neural activity eventually spreads across the whole assembly,
saturating the network.

It would not be until 1957 when Frank Rosenblatt \textemdash who had
previously read \textit{The Organization of Behavior} and sought to
find a more ``model-friendly'' version of Hebb's assembly\textemdash{}
came with a solution: the Perceptron, the first example of a Feed
Forward Neural Network (FFNN) \citep{rosenblatt_perceptron_1958}.
Dismissing the idea of a homogeneous mass of cells, Rosenblatt introduced
three different types of units within the network, which would correspond
today to what is usually known as input, hidden and output layers
in a FFNN. Mathematically, the output of the perceptron is computed
as: 
\begin{equation}
f(x)=\begin{cases}
1 & if\,w\cdot x+b>0\\

0 & otherwise
\end{cases}
\end{equation}

where $w\cdot x$ is the dot product of the input $x$ with the weight
vector $w$ and $b$ is a bias term that acts like a moving threshold.
In modern FFNNs, the step function is usually substituted by a non-linearity
$\varphi\left(w\cdot x+b\right)$ which receives the name of \textit{activation
function}. Being computationally more applicable than the original
ideas of Hebb, Rosenblatt paved the way that would progressively detach
ML from its biological inspiration.

Despite the initial excitement, in 1969 Marvin Minsky and Seymour
Papert proved that perceptrons could only be trained to recognize
linearly separable patterns \citep{minsky_perceptrons:_1969}. The
authors already foresaw the need for Multilayer Perceptrons (MLP)
to tackle non-linear classification problems, but the lack of suitable
learning algorithms lead to the first of the AI winters \citep{kurenkov_brief_2015},
with neural network research stagnating for many years. The thaw would
not arrive until 1974 with the advent of today's widely known backpropagation
algorithms \citep{linnainmaa_representation_1970,werbos_new_1974}.
Understood as a supervised learning method in multilayer networks,
backpropagation aims at adjusting the internal weights in each layer
to minimize the error or loss function at the output using a gradient-descent
approach. Despite their success in tasks as diverse as speech recognition,
natural language processing, medical image analysis or board game
programs; backpropagation methods lack of a corresponding biological representation.
Instead, ANN that aim to resemble the biology behind the operation of the human brain
ought to include neurons that send feedback signals to each other. This
is the idea behind a Recurrent Neural Network (RNN). Whereas FFNNs
are able to approximate a mathematical function, RNNs can approximate
dynamical systems \textemdash i.e. functions with an added time component\textemdash{}
so that the same input can result in a different output at different
time steps \citep{grezes_reservoir_2014}.

It is within this context when two fundamentally new approaches to
RNNs appeared independently: the Echo State Network (ESN) \citep{jaeger_echo_2001}
and the Liquid State Machine (LSM) \citep{maass_real-time_2002},
both constituting trailblazing models of what today is known as the
Reservoir Computing (RC) paradigm. These models are particularly fast
and computationally much less expensive since training happens only
at the output layer through the adjustment of the readout weights.
Although very flexible, this approach also leaves the open question
of how to choose the reservoir connectivity to maximize performance.
While most reservoir computing approaches consider a reservoir with fixed internal connection weights, plasticity was rediscovered as an unsupervised,
biologically inspired adaptation to implement an adaptive reservoir. It appeared
first as a type of Hebbian synaptic plasticity to modify the reservoir
weights \citep{babinec_improving_2007}, but soon the ideas of nonsynaptic
plasticity that inspired the first Intrinsic Plasticity (IP) rule
\citep{triesch_gradient_2005} were also implemented in an Echo State
Network \citep{schrauwen_improving_2008}. After that, many different
models of plasticity rules have been implemented in RC networks with promising results \citep{steil_online_2007,yusoff_modeling_2016,wang_echo_2019}.
Today, the fact that biologically meaningful learning algorithms have
a place in these models, together with recent discoveries suggesting
that biological neural networks display RCs' properties \citep{ju_spatiotemporal_2015,enel_reservoir_2016},
make reservoir computing a field of machine learning in continuous
growth.

Echo State Networks have been shown to successfully perform in a wide number of tasks,
ranging from speech recognition \citep{skowronski_automatic_2007},
channel equalization \citep{jaeger_harnessing_2004}, or robot control \citep{hertzberg_learning_2002},
to stock data mining \citep{lin_application_2008}. Here, we will focus in the challenging problem of chaotic time
series forecasting. This type of task has been addressed for a large
number of different time series \citep{babinec_gating_2009,lin_short-term_2009,yusoff_modeling_2016,wang_time_2019}, and ESNs implementing plasticity rules to improve time series forecasting
have been treated before in \citep{yusoff_modeling_2016,wang_time_2019,babinec_improving_2007}.
Nevertheless, in this paper we will move away from the finest-performance
approach, focusing instead in understanding how unsupervised learning through plasticity rules affects the ESN architecture in a way that boosts its performance.

The paper is structured as follows. The Methods section includes the standard definition of the ESN, the models considered for synaptic and intrinsic forms of plasticity, as well as the measures that will be employed for performance characterization. We consider the so-called anti-Hebbian types of learning rules for synaptic plasticity, which in our case means that neurons' activity at subsequent times tend to become decorrelated. As for intrinsic plasticity, we modify the parameters of the response functions of individual neurons to accommodate to a target Gaussian distribution function. In the Results section, we find that the best performance is usually obtained by employing a combination of both, synaptic and intrinsic plasticity, thus revealing the emergence of synergistic effects. Finally, we discuss the influence of the plasticity rules on the dynamical response of the individual neurons as well as on the global activity of the reservoir.

\section{Methods}

\subsection{The ESN model: architecture, training and testing.}

\begin{figure}
\begin{centering}
\includegraphics[width=\columnwidth]{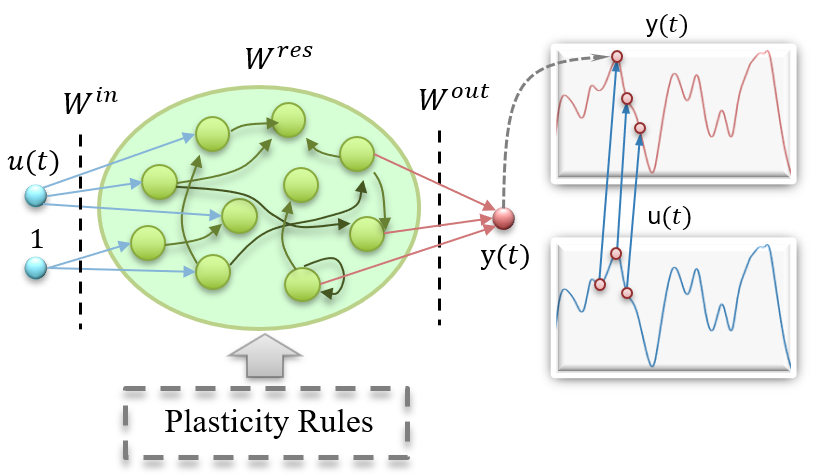} 
\par\end{centering}
\caption{Architecture and functioning of a basic Echo State Network for a one-step-ahead chaotic time series prediction task.\label{fig:Architecture-of-a}}
\end{figure}

The basic architecture of an ESN model is made of three layers: an input layer, a hidden layer or reservoir, and an output layer. Fig. \ref{fig:Architecture-of-a} illustrates the ESN architecture, where we already particularized the more general concept for
two input units \textemdash one feeding a point of the series at each discrete time step and a second one acting as a bias\textemdash{} and one output neuron. 

In Fig. \ref{fig:Architecture-of-a}, points $u(t)\in\mathbb{R}$ in the temporal series are fed as input
after being multiplied by a weight matrix $W^{in}\in\mathbb{R}^{N_{x}\times2}$.
The internal connections between neurons in the reservoir are defined by $W^{res}\in\mathbb{R}^{N_{x}\times N_{x}}$, where $N_{x}$ is
the number of neurons in the reservoir. The states of
the neurons in the reservoir produce the final output after multiplying by an output weight matrix $W^{out}\in\mathbb{R}^{1\times N_{x}}$.
Thus, the network dynamics for the reservoir and readout states are given by: 
\begin{align}
\mathbf{x}(t)&=tanh(\epsilon W^{in}[1;u(t)]+W^{res}\mathbf{x}(t-1)),\label{eq:States_Update}\\
y(t)&=W^{out}\mathbf{x}(t),\label{eq:Output_Eq}
\end{align}
where $\epsilon$ is the input scaling, and $W^{in}$ and $W^{res}$ are often randomly initialized. Here we chose the hyperbolic tangent as our activation function, but it could be in general any nonlinear function. Using a supervised learning scheme, the goal is to generate an output $y(t)\in\mathbb{R}$
that not only matches as closely as possible the desired target $y^{target}(t)\in\mathbb{R}$ but can also generalize to unseen data. Because large output weights are commonly associated to overfitting of the training data \citep{lukosevicius_practical_2012}, it is a common practice to keep their values low by adding a regularization term to the error in the target reconstruction. Although several regularization methods have been proposed  \citep{jaeger_echo_2001,reinhart_constrained_2010,reinhart_reservoir_2011},
here we use the Ridge regression method, for which the error is defined
as: 
\begin{equation}
E_{ridge}=\dfrac{1}{T}\sum_{t=1}^{T}\left(y^{target}(t)-y(t)\right)^{2}+\beta\left\Vert W^{out}\right\Vert ^{2},\label{eq:MSE_Error}
\end{equation}
where $\left\Vert \text{\ensuremath{\cdot}}\right\Vert $stands
for the Euclidian norm, $\beta$ is the regularization coefficient
and $T$ is the total number of points in the training set. Notice
that choosing $\beta=0$ removes the regularization, turning the ridge
regression into a generalized linear regression problem. After training, the expression for the optimal readout weights $W^{out}_{opt}$ can be easily  obtained \textemdash{minimizing the above error}\textemdash{} as:
\begin{equation}
W^{out}=Y^{target}X^{T}\left(XX^{T}+\beta I\right)^{-1},\label{eq:Ridge}
\end{equation}
where $I$ is the identity matrix, $Y^{target}\in\mathbb{R}^{1\times T}$ contains all output targets $y^{target}(t)$, and ${X\in\mathbb{R}^{\left(1+N_{u}+N_{x}\right)\times T}}$ consists of all concatenated vectors $[1;u(t);x(t)]$. It is worth noticing that the standard training in ESNs focuses on the optimization of the readout weights, $W^{out}$, 
but does not modify the initial reservoir, which is usually considered randomly connected. A natural step forward is then to
optimize the weights of the reservoir connections, $W^{res}$, or the excitability of the neurons according to the inputs reaching the reservoir, that is, to introduce rules of neuronal plasticity.

\subsection{Neuronal plasticity: biological and artificial implementations.}

The term plasticity has been used in brain science for well over a
century to refer to the suspected changes in neural organization that
may account for various forms of behavioral changes, either short-
or long-lasting \citep{berlucchi_neuronal_2008}. From a biological
point of view, mechanisms of plasticity in the brain can be grouped
into two large categories: synaptic and nonsynaptic. Synaptic plasticity
deals directly with the strength of the connection between neurons,
which is linked to the amount of neurotransmitter released from the
presynaptic neuron and the response generated in the postsynaptic
channels \citep{abbott_synaptic_2000,song_regulation_2002,gerrow_synaptic_2010}.
Nonsynaptic plasticity, instead, involves modification of the intrinsic
excitability of the neuron itself, operating through structural changes
that usually affect voltage-dependent membrane conductances in the
axon, dendrites or soma \cite{turrigiano1999homeostatic,marder2006variability}.

Likewise, but now from the perspective of RC, plasticity rules aim
to modify either the weights of the connections $W^{res}$ (synaptic
plasticity) or the excitability of the reservoir units (nonsynaptic
plasticity) based on the activity stimulated by the input. In this manner, the information carried by the input signal is partly embedded in the reservoir. Although non-Hebbian forms of synaptic plasticity have been found empirically \citep{alonso_postsynaptic_1990,kato_non-hebbian_2009},
most rules modifying the synaptic strength among neurons fall into
the category of Hebbian learning. The Hebbian rule, as originally
proposed by Hebb \citep{hebb_organization_1949}, can be described
mathematically as a change in the synaptic strength between two neurons that is proportional to the product of the pre and post-synaptic activities at time $t$: 
\begin{equation}
w_{kj}(t+1)=w_{kj}(t)+\eta x_{j}(t)x_{k}(t+1),\label{eq:hebb}
\end{equation}
where $w_{kj}$ is the weight of a synapse connecting neurons
k and j \textemdash with j triggering the activity of k\textemdash ;
$x_{j}(t)$ and $x_{k}(t+1)$ represent the activity of the pre and post-synaptic neurons, $\eta$ is a parameter accounting for the learning rate, and all weights in the reservoir are updated in parallel at each discrete time step. Notice that we refer to matrices using capital letters and denote with small letters their elements. 
The growth of the weights in the direction of the correlations between pre and post-synaptic units has an obvious flaw: as the connections get stronger following Hebb's postulate, activity will eventually spread and increase uncontrollably throughout the network. To avoid this, one possibility is to normalize
the weights arriving to each post-synaptic neuron $x_{k}$, so that
$\sqrt{\sum_{j}w_{kj}^{2}}=1$. We can then rewrite the update rule in  Eq. \ref{eq:hebb} as: 
\begin{equation}
w_{kj}(t+1)=\dfrac{w_{kj}(t)+\eta x_{k}(t+1)x_{j}(t)}{\sqrt{\sum_{j}\left(w_{kj}(t)+\eta x_{k}(t+1)x_{j}(t)\right)^{2}}}.\label{eq:anti-Hebb}
\end{equation}
Note that Eq. \ref{eq:anti-Hebb} is non-local (NL), meaning that a modification in a given weight $w_{kj}$ also depends on other neurons in addition to the connected neurons $k$ and $j$.  Finally, assuming a small learning rate $\eta$ and linear activation functions in the absence of external inputs, Oja derived a local approximation to Eq. \ref{eq:anti-Hebb}, known today as Oja's rule \citep{oja_simplified_1982}:
\begin{equation}
w_{kj}(t+1)=w_{kj}(t)+\eta x_{k}(t+1)\left(x_{j}(t)+x_{k}(t+1)w_{kj}(t)\right).
\label{eq:oja}
\end{equation}

It has been suggested that a change in the sign of Hebbian plasticity rules may be advantageous in making an effective use of the dynamic range of cortical neurons \cite{barlow1989adaptation}, while also promoting decorrelation between the activity induced by different inputs.

Therefore, in this paper, we will work with such so-called anti-Hebbian learning rules, which are obtained simply by changing the signs of the weight update in Eqs. \ref{eq:hebb}, \ref{eq:anti-Hebb},  and \ref{eq:oja}. The precise writing of the anti-Hebbian learning rules used here and a complete derivation of the anti-Oja rule can be found in App. \subsecref{S.I.-2.-Derivation}. For the sake of clarity we stress that, from a practical point of view, the synaptic strengths $w_{kj}$ updated with the plastic rules correspond to the reservoir weights $w_{kj}^{res}$ of our ESN models.

Although there are examples of the anti-Oja rule applied to ESNs with nonlinear activation functions  \citep{babinec_improving_2007,yusoff_modeling_2016},
Eq. \ref{eq:oja} is strictly valid only when the state of the
post-synaptic neurons is a linear combination of the pre-synaptic
states in the form $x_{k}(t+1)=\sum_{j}w_{kj}x_{j}(t)$, which is no longer true in the presence of nonlinear neurons. In order to evaluate the influence of the local approximation derived by Oja, we will compare the performance obtained by using Eq. \ref{eq:anti-Hebb} (with the minus sign, see Eq. \ref{eq:NL_Hebb}) and the one obtained by using Eq. \ref{eq:oja} (with the minus sign, see Eq. \ref{eq:antiOja}). 
We will show in the Results section that the NL anti-Hebbian rule outperforms the anti-Oja rule in chaotic time series prediction tasks.

We now consider the intrinsic plasticity (IP), which adjusts the neurons'
internal excitability instead of the individual synapses. Based on
the idea that every single neuron intends to maximize its information
transmission while minimizing its energy consumption, Jochen Triesch
proposed a mathematical learning rule that leads to maximum entropy
distributions for the neurons output with certain fixed moments \citep{triesch_gradient_2005}.
Although the original derivation of Triesch applied to Fermi activation functions
and exponential desired distributions, soon Schrauwen et al. \citep{schrauwen_improving_2008}
extended the rule to account for neurons with hyperbolic tangent functions. In this case, each neuron updates its state through the following expression: 
\begin{equation}
x_{k}(t)=tanh(a_{k}z_{k}(t)+b_{k})
\end{equation}
where $a_{k}$ and $b_{k}$ are the gain and bias of the post-synaptic
neuron, and $z_{k}(t)=\epsilon w_{k}^{in}[1;u(t)]+\sum_{j}w_{ij}x_{j}(t-1)$
is the total arriving input. The minimization of the Kullback-Leibler divergence with respect to a desired Gaussian output distribution with a given mean and variance leads to the following online learning rules for the gain and bias: 
\begin{equation}
\Delta b_{k}=-\eta\left(-\dfrac{\mu}{\sigma^{2}}+\dfrac{x_{k}(t)}{\sigma^{2}}\left(2\sigma^{2}+1-x_{k}(t){}^{2}+\mu x_{k}(t)\right)\right),
\label{eq:bk}
\end{equation}
\begin{equation}
\Delta a_{k}=\dfrac{\eta}{a_{k}}+\Delta b_{k}z_{k}(t),
\label{eq:ak}
\end{equation}
where $\eta$ is the learning rule and $\mu$ and $\sigma$ the mean
and standard deviation of the targeted distribution, respectively. 

Finally, we will also consider the combination of two of the above rules \textemdash the NL anti-Hebbian
and IP algorithms\textemdash{} to assess the performance of an ESN
when these two types of plasticity act in a synergistic manner. For this combination,
there are three natural ways in which the training can be carried
out: $\sl i)$ applying both rules simultaneously to update the intrinsic
parameters and connections weights after each input; $\sl ii)$ modifying
first the connections through the synaptic plasticity and then applying
the IP rule; or $\sl iii)$ conversely, changing first the intrinsic
plasticity of the neurons and then the synapses strength among them.
From all the alternatives, the application of the NL anti-Hebbian
rule through the whole training set followed by the application of the IP rule through the same training set yielded the best performance, and is therefore used in
the forthcoming results section. Computational models combining the effect
of synaptic and non-synaptic plasticity have been previously suggested
in the literature for simple model neurons \citep{triesch_synergies_2007},
FFNNs \citep{li_synergies_2013} and RNNs \citep{li_synergies_2013,janowitz_excitability_2006,aswolinskiy_rm-sorn:_2015}.
However, we find that a simple combination of two standard plasticity
rules can ease the tractability of the results, while allowing fairer
comparisons against the other plasticity models.  

\subsection{Prediction of a chaotic time series.}

The task at hand consists on the prediction of the points continuing
a Mackey-Glass series, a classical benchmarking dataset generated
from a time-delay differential equation (see App. \subsecref{S.I.-1.-Mackey-Glass}
for details on the generation of the dataset). Since this series exhibits
a chaotic behavior when the time delay $\tau>16.8$, we construct
two different sets: one with $\tau=17$ (MG-17), often used as an
example of mildly chaotic series; and a second one with  $\tau=30$ (MG-30) that presents stronger chaotic behavior. To assess its performance, we initially feed the ESN with the last input of the training set, $u(T)$, then run the network for a number $F$ of steps using the predicted output at time $t$ as the next input at time $t+1$ (i.e. $u(t+1)=y(t)$). In this manner, the testing phase is done in the so-called autonomous or generative mode with output feedback. To quantify the error for this task, we use two different quantities: 
\begin{itemize}
\item The root mean square error ($RMSE$) over the predicted continuation of the series: 
\begin{equation}
RMSE=\sqrt{\dfrac{1}{F}\sum_{t=0}^{F}\left(y(t)-y_{target}(t)\right)^{2}}
\end{equation}
\item The furthest predicted point (FPP): this is the furthest point up to which the trained ESN is able to continue
the series without significantly deviating from the original one.
The tolerance for significant deviation is taken as $\varepsilon=0.02$,~which
represents approximately 2\% of the maximum distance between any two
points in the original MG-17 and MG-30 series. 
\end{itemize}

\subsection{Memory capacity task.}

The task of memory capacity (MC) is based on the network's ability
to retrieve past information from the reservoir using the linear combinations
of reservoir unit activations. To assess the ability of each ESN model
to restore previous inputs fed into the network, we compute the (short-term)
MC as introduced by Jaeger in \citep{jaeger_short_2001}: 
\begin{equation}
MC=\sum_{d=1}^{\infty}MC_{d}=\sum_{d=1}^{\infty}\dfrac{cov^{2}(u(t-d),y_{d}(t))}{var(u(t))\cdot var(y_{d}(t))}
\end{equation}
where $cov$ and $var$ denote covariance and variance, respectively.
In the above expression, $u(t-d)$ is the input presented $d$ steps
before the current input $u(t)$, and $y_{d}(t)=W_{d}^{out}X=\tilde{u}(t-d)$
is its reconstruction at the output unit $d$ with trained output weights $W_{d}^{out}$. A value $MC_{d}\sim1$
means that the system is able to accurately reconstruct the input
fed to the network $d$ steps ago. Thus, the sum of all $MC_{d}$
represents an estimation of the number of past inputs the ESN is able
to recall. Although the sum runs to infinity in the original definition
\textemdash accounting for the complete past of the input\textemdash{}
in practice the data fed is finite and it will suffice with setting $d_{max}=L$,
with $L$ being the number of output units of the ESN. Each of the
$L$ output units is independently trained to approximate past inputs
with a different value of $d$. A theoretical limit for the memory
capacity was derived in \citep{jaeger_short_2001} to be $MC_{max}\approx N-1$,
with $N$ the number of reservoir neurons.

\section{Results}

\subsection{Hyper-parameter optimization.\label{sec:hyper}}

One of the biggest drawbacks of Echo State Networks is their high
sensitivity to hyper-parameters choice (see \citep{lukosevicius_practical_2012}
for a detailed review on their effects over the network performance).
In this work, we focus on tuning four hyper-parameters to improve the performance of each ESN model: the reservoir size or number of neurons in the reservoir $N$, the input scaling $\varepsilon$, the
spectral radius $\rho$ of the reservoir's weight matrix (i.e. the
maximum absolute eigenvalue of $W^{res}$) and the regularization parameter $\beta$
in the ridge regression. Weights in the reservoir and input
layers are initialized randomly according to a uniform distribution
between -1 and 1. 
Sparseness in the reservoir matrix is set to 90\%,
meaning that only 10\% of all connections have initially a non-zero
value. When incorporating plasticity rules, an extra tunable hyper-parameter
$\eta$ describing the learning rate in the update rules is included.
When IP is implemented, we find that best results are obtained when
using $\mu=0$ and $\sigma=0.5$ as the mean and variance of the targeted
distribution for the neuron states. For the sake of comparison between different ESN models, we choose initially a common non-optimal, but generally well-performing set of hyper-parameters $\{\rho=0.95,\,\varepsilon=1,\,\beta=10^{-7}\}$ for all of them, with $ N=300,\,\eta=10^{-6}$ for the MG-17 series prediction and $N=600,\,\eta=10^{-7} $ for the MG-30.

\subsection{Performance in prediction tasks.}

\begin{figure}
\begin{centering}
\includegraphics[width=\columnwidth]{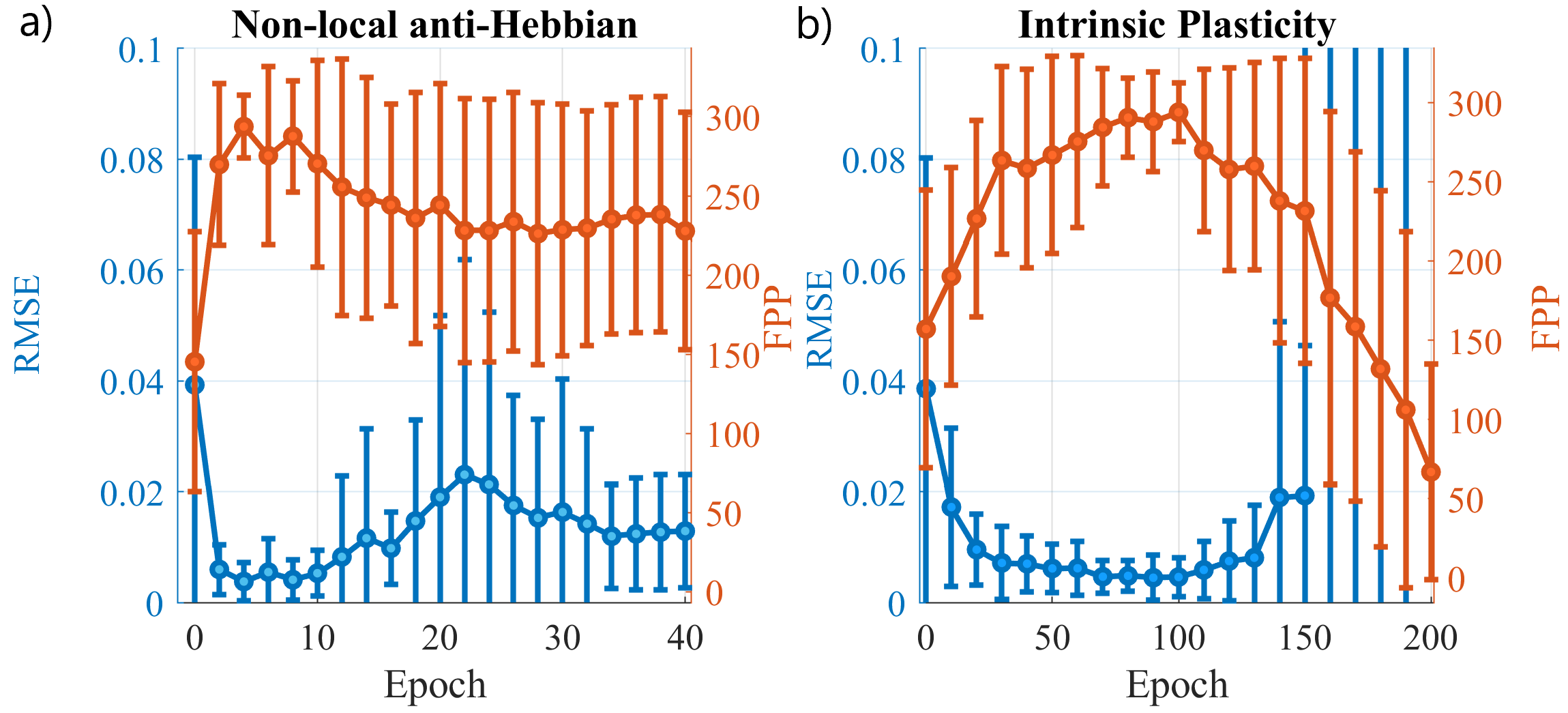} 
\par\end{centering}
\caption{Evolution of the total RMSE over $F=300$ predicted points
and FPP for the MG-17 chaotic time series prediction task as a function of the number of training episodes for different plasticity rules.
At each epoch, averages were computed over 20 independent realizations
of a 300 neurons ESN.\label{fig:Evolution-of-the}}
\end{figure}

In order to compare the influence of the different plasticity rules, we first estimate the number of plasticity training epochs that optimizes the performance for each model. 
We note that the neuronal plasticity rules are only active in the unsupervised learning procedure but not during the prediction, as detailed in App. A. The unsupervised learning can last for several epochs, with each epoch containing $T=4000$ points of the time series for the MG-17 task. Once the plastic unsupervised learning has finished, $W^{out}$ is computed in a supervised fashion after letting the reservoir evolve for an additional $T=4000$ steps.
Fig. \ref{fig:Evolution-of-the} shows the evolution of the RMSE and FPP for the non-local anti-Hebbian and IP rules.
In this figure, the optimal number of epochs can be easily found as the point in which the RMSE (FPP) presents a global minimum (maximum). As we see, the performance gets worse as the ESNs with plasticity are over-trained.
We will focus on understanding the role of plasticity rules in Sec. \ref{subsec:Under}.

In Table \ref{tab:RMSE,-NRMSE84-and} we show the results obtained for the anti-Oja, NL anti-Hebbian and IP rules when each model is trained optimally (i.e. for the optimal number of epochs). The number of epochs for the Anti-Oja, NL anti-Hebbian, and IP rules are 10, 8, and 100, respectively. For the sake of comparison, we also include in Table \ref{tab:RMSE,-NRMSE84-and} the results for a non-plastic ESN with the hyper-parameters mentioned at the end of Sec. \ref{sec:hyper}.
It can be observed that the implementation of plasticity rules reduces the average prediction error and its uncertainty, specially in the highly chaotic series MG-30, while keeping consecutively predicted points close to the original test set for a longer time.  
For the chosen set of hyper-parameters, the non-plastic ESN is only able to properly predict autonomously the initial points of the MG-30 chaotic time series. 
Thus, the corresponding error computed over $F=100$ points of the testing time-series is very large.

When comparing different plastic rules in Table \ref{tab:RMSE,-NRMSE84-and}, we find that the NL anti-Hebbian and IP rules yield a better prediction than the anti-Oja one. In addition, we find that the combination of NL anti-Hebbian and IP reaches the lowest RMSE and largest FPP, thus providing "better and further" predictions.

\begin{table*}
\begin{centering}
\caption{RMSE and Furthest Predicted Point (FPP) on the MG-17 and MG-30 prediction tasks for different implementations of plasticity rules. The RMSE was calculated using $F$ steps of the predicted time-series, with $F=300$ for the MG-17 and $F=100$ for the MG-30. Averages were computed over 20 independent realizations. \label{tab:RMSE,-NRMSE84-and} }
\begin{tabular}{ccccccc}
\toprule 
\multicolumn{2}{c}{} & {\footnotesize{}{}Non-Plastic}  & {\footnotesize{}{}Anti-Oja}  & {\footnotesize{}{}NL anti-Hebbian}  & {\footnotesize{}{}IP}  & {\footnotesize{}{}NL anti-Hebbian + IP}\tabularnewline
\midrule
\midrule 
\multirow{2}{*}{{\footnotesize{}{}MG17}}  & {\footnotesize{}{}RMSE}  & {\footnotesize{}{}$0.05\pm0.05$}  & {\footnotesize{}{}$0.02\pm0.03$}  & {\footnotesize{}{}$0.004\pm0.004$}  & {\footnotesize{}{}$0.004\pm0.002$}  & {\footnotesize{}{}$0.003\pm0.002$}\tabularnewline
 & {\footnotesize{}{}FPP}  & {\footnotesize{}{}$136\pm73$}  & {\footnotesize{}{}$208\pm89$}  & {\footnotesize{}{}$288\pm35$}  & {\footnotesize{}{}$289\pm33$}  & {\footnotesize{}{}$299\pm2$}\tabularnewline
\midrule 
\multirow{2}{*}{{\footnotesize{}{}MG30}}  & {\footnotesize{}{}RMSE}  & {\footnotesize{}{}$1\pm3$}  & {\footnotesize{}{}$0.03\pm0.02$}  & {\footnotesize{}{}$0.011\pm0.010$}  & {\footnotesize{}{}$0.018\pm0.011$}  & {\footnotesize{}{}$0.011\pm0.011$}\tabularnewline
 & {\footnotesize{}{}FPP}  & {\footnotesize{}{}$7\pm4$}  & {\footnotesize{}{}$53\pm31$}  & {\footnotesize{}{}$82\pm29$}  & {\footnotesize{}{}$51\pm37$}  & {\footnotesize{}{}$85\pm32$} \tabularnewline
\bottomrule
\end{tabular}
\end{centering}

\end{table*}

\subsection{Performance in memory capacity task.}

\begin{figure}
\begin{centering}
\includegraphics[width=\columnwidth]{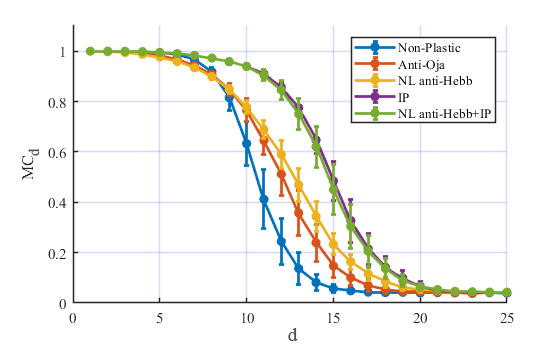} 
\par\end{centering}
\caption{Memory curves, $MC_d$, as a function of the delay $d$ for the different models of ESNs studied. Averages
of the values were taken over 20 independent realizations of each
model.\label{fig:Forgetting-curves-for} }
\end{figure}

We construct single input node ESNs with $N=150$ reservoir neurons and $L=300$ output nodes, such that $MC_d$ is computed up to a delay ${d_{max}=L}$.
For this task, we feed the network with a random time series of $T=4000$ points, drawn from a uniform probability distribution in the interval {[}-1,1{]}. Fig. \ref{fig:Forgetting-curves-for} shows the memory curves for an ESN before and after implementation of the different plasticity rules. 
Again, we notice how models with implemented plasticity outperform the original non-plastic ESN, with the memory decaying faster in the latter case.
In Table \ref{tab:Memory-Capacity-for}, we present the estimated MC computed for the plastic and non-plastic versions of the ESN.
Here, we find that the IP rule and the combination of NL anti-Hebbian and IP yield the largest memory capacities.
These results are in agreement with the average values presented in \citep{boedecker_information_2011}, where the maximum memory was observed at the edge of stability for a random recurrent neural network.

In the next section, we are going to explore in more detail the properties of the plastic ESNs.

\bigskip{}
 
\begin{table*}
\begin{centering}
\caption{Memory Capacity for an ESN with 150 nodes and 300 output neurons with
and without implementation of the different plastic rules.\label{tab:Memory-Capacity-for}}
\begin{tabular}{cccccc}
\toprule 
 & Non-Plastic  & Anti-Oja  & NL anti-Hebb  & IP  & NL anti-Hebb + IP\tabularnewline
\midrule 
MC  & $21.6\pm0.4$  & $23.0\pm0.5$  & $23.6\pm0.3$  & $25.7\pm0.5$  & $25.7\pm0.5$\tabularnewline
\bottomrule
\end{tabular}
\par\end{centering}
\end{table*}

\subsection{Understanding the effects of plasticity.\label{subsec:Under}}

\subsubsection*{Influence of plasticity rules on the reservoir dynamics.}
\begin{figure}
\begin{centering}
\includegraphics[width=\columnwidth]{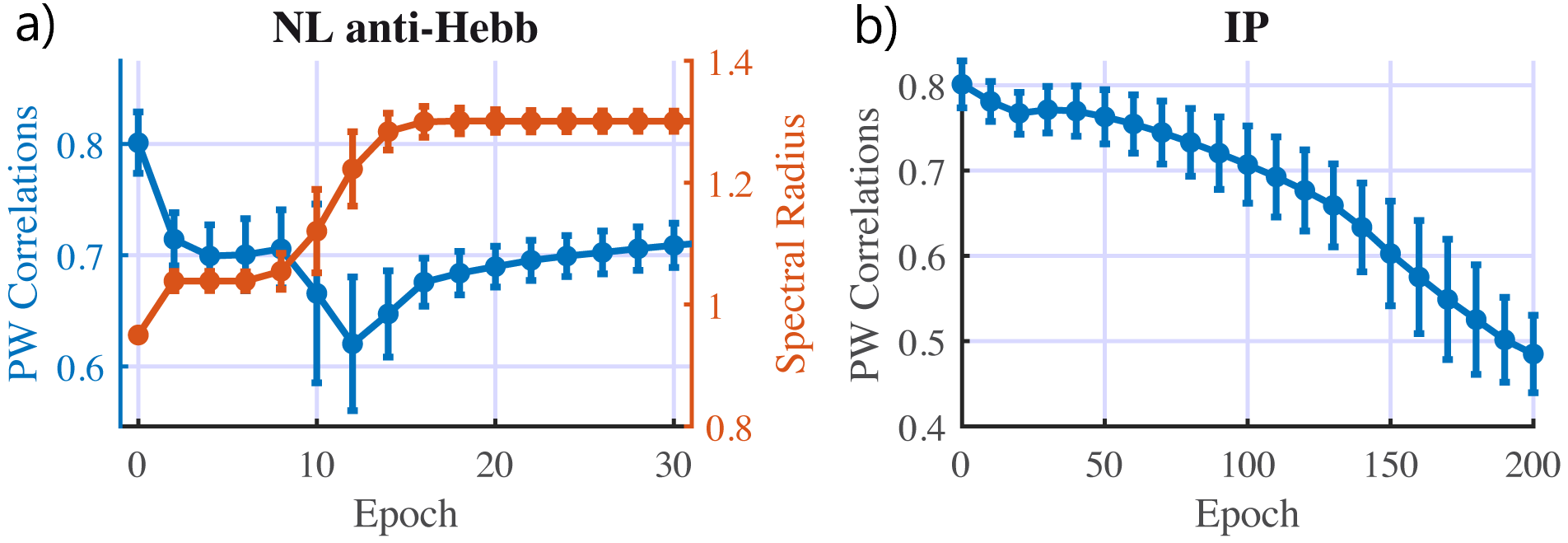} 
\par\end{centering}
\caption{Evolution of Pearson correlation coefficient as a function of the training epochs for the a) NL anti-Hebbian and b) IP rules. For the NL anti-Hebbian, the evolution of the spectral radius of $W^{res}$ is also shown.\label{fig:Evolution-of-the-5}}
\end{figure}

To analyze the effects of plasticity over the ESN performance, we focus now on the MG-17 prediction task, casting our attention into the dependence of the performance on the number of training epochs. 
As mentioned above, 
Fig. \ref{fig:Evolution-of-the} shows that the measures of performance exhibit absolute extrema (minimum of the errors, maximum of the number of predicted points), which are followed by a worsening of the predictions as the number of epochs increase.
In order to understand this behavior, we studied quantities related to the reservoir dynamics as the plasticity training advanced.
In Fig. \ref{fig:Evolution-of-the-5}, we show the average absolute Pearson correlation coefficient among reservoir states at consecutive times, as defined in App. \subsecref{S.I.-3.-Measures}. In addition, and for the case of synaptic plasticity only, we present
the spectral radius of the reservoir matrix (which does not change in the IP rule) as the non-supervised plasticity training evolves .

Focusing first on the NL anti-Hebbian rule, we observe in Fig. \ref{fig:Evolution-of-the}a) that the prediction error increases significantly beyond 10 training epochs. This fact could be attributed in first place to the associated increase of the reservoir matrix spectral radius, as seen in Fig. \ref{fig:Evolution-of-the-5}a). A maximum absolute eigenvalue exceeding unity has often been regarded as a source of instability in ESNs due to the loss of the ``echo state property'', a mathematical condition ensuring that the effect of the initial conditions dies out asymptotically with time \citep{jaeger_echo_2001,lukosevicius_reservoir_2009,lukosevicius_practical_2012}. Nevertheless, subsequent studies proved that the echo state property can be actually maintained over a unitary spectral radius, depending on the input fed to the reservoir \citep{yildiz_re-visiting_2012,manjunath_echo_2013}, which could be the reason why we find optimal performance slightly above $\rho=1$.

The results presented here seem to agree with the results presented in \citep{boedecker_information_2011}, where it was suggested that information transfer and storage in ESNs are maximized at the edge between a stable and an unstable (chaotic) dynamical regime. In our case the ESN becomes unstable (periodic) for $\rho\sim1.2$ and we find an associated decrease in the memory capacity of the ESN. A chaotic dynamic inside the reservoir is not observed in our numerical simulations.

Additionally, we find that the increase in the prediction error coincides with a sharp decrease in the consecutive-time pair-wise absolute correlations as shown in Figs. \ref{fig:Evolution-of-the} and \ref{fig:Evolution-of-the-5}. This decrease in the correlations, which was to be expected in any anti-Hebbian type of rule \textemdash by their very own definition \textemdash{} occurs also along the training of the IP rule. This remarkable common trend  hints at the possibility that, to some extent, decorrelation inside the reservoir could indeed enhance the network computational capability. However, an over-training of the plasticity rules yields an error increase.

\begin{figure}
\begin{centering}
\includegraphics[width=\columnwidth]{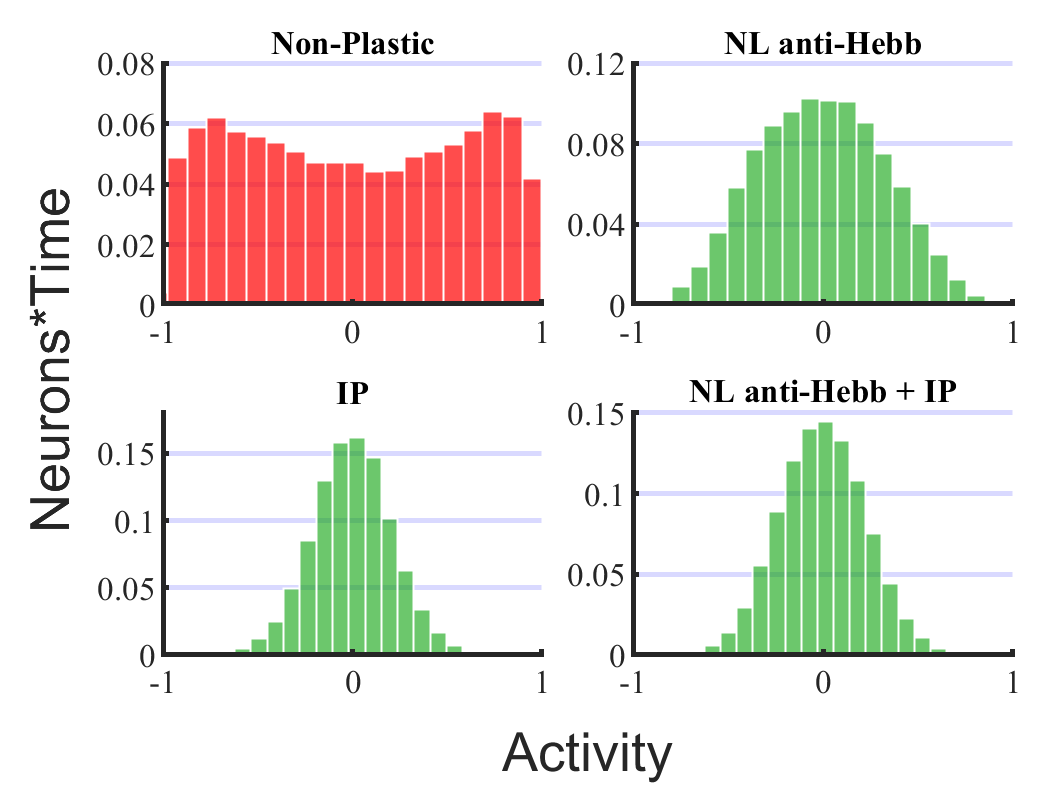} 
\par\end{centering}
\caption{Distribution of reservoir states after training for the explored ESN models. Histograms were constructed averaging over 20 different realizations of the corresponding
ESNs. \label{fig:Distribution-of-states}}
\end{figure}

To further evaluate the effects of plasticity over the dynamics of the reservoir, we also analyzed the distribution of the reservoir states $\mathbf{x}(t)$ before and after implementation of each rule. Keeping the values of all the states
at each input $u_{t}$ of the training sequence, then averaging the
resulting matrix over 20 realizations with different training sets,
an ``average'' histogram describing the distribution
of the states is presented in Fig. \ref{fig:Distribution-of-states}.
It can be observed that the application of the plasticity rules changes the distribution of the reservoir states from a rather uniform shape to a unimodal one.
The initial uniform shape is given by our choice of uniform input weights and the given input scaling.
As expected from the mathematical formulation of the IP rule, the distribution of the states after its implementation approaches that of a Gaussian centered around zero.
Remarkably, the application of synaptic plasticity also shapes the initial distribution to a unimodal one peaking around zero.
The observed distribution of reservoir states after the application of plasticity rules entails that the individual reservoir neurons tend to avoid operating at the saturation of the  $tanh$ non-linearity.

\subsubsection*{Influence of plasticity rules on the neuron dynamics.}
So far, we have focused on understanding the effects of plasticity at the network level, but nothing has been said about the way each individual neuron ``sees'' or ``reacts'' to
the input after implementation of the plastic rules. To shed some
light into this question, we define the effective input $\widetilde{u}_{n}(t)$
of a neuron $n$ at time $t$ as the sum of the input and bias unit once filtered through the input mask, $\widetilde{u}_{n}(t)=w_{0n}^{in}+w_{1n}^{in}\cdot u_{n}(t)$.
In this way, the state update equation for a single neuron (with no IP implemented) can be rewritten as:

\begin{equation}
x_{n}(t+1)=tanh\left(\epsilon\widetilde{u}_{n}(t)+\sum_{j}w_{nj}^{res}x_{j}(t)\right)\label{eq:update_feed}
\end{equation}

\begin{figure}
\begin{centering}
\includegraphics[width=\columnwidth]{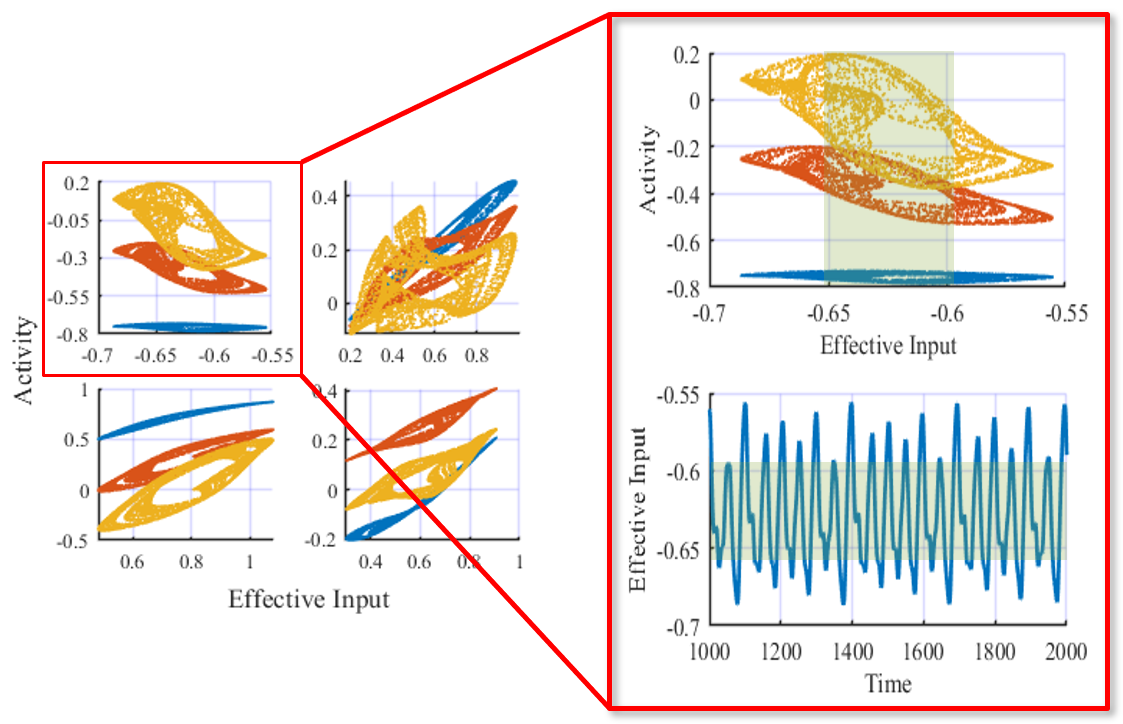} 
\par\end{centering}
\caption{Activity of 4 different neurons as a function of the effective input in a non-plastic ESN (blue) and in the same reservoir after training it with NL anti-Hebbian
rule for 8 epochs (red) and IP rule for 100 epochs (yellow). On
the right side we zoom in one of the neurons, plotting also the
evolution of the effective input over a section of the training. We highlight in green the range of inputs for which the activity broadens more notably with respect to the non-plastic case, coinciding with one of the most variable parts of the input. \label{fig:Activity-vs-filtered-1}}
\end{figure}

In Fig. \ref{fig:Activity-vs-filtered-1}, we plot the response of 4 different neurons to this effective input before (blue dots) and after (red and yellow dots) the implementation of the non-local anti-Hebbian and IP rules. On the right side we have zoomed in on one of these neurons and plotted also 1000 points of the effective input $\widetilde{u}_{n}(t)$ that arrives to it. It can be clearly seen that plasticity has the effect of widening the activity range of the neurons, specially
on those areas \textemdash as highlighted in green\textemdash{} in
which the same point may lead to very different continuations of the
series depending on its past. To quantify this widening, we measured the average area of the reservoir neurons' activity phase space before and after implementation of the plasticity rules.
The results for these average phase space areas, as presented in Table \ref{tab:Average-neuron-phase}, back up the aforementioned expansion, which is specially significant in the case of the IP rule.
\begin{table*}
\begin{centering}
\caption{Average neuron phase space area for plastic and non-plastic implementations of a 300 neurons ESN. The error is given as the standard deviation over all neurons. \label{tab:Average-neuron-phase}}
\begin{tabular}{ccccc}
\toprule 
Non-Plastic  & Anti-Oja  & NL anti-Hebb  & IP  & NL anti-Hebb + IP\tabularnewline
\midrule 
$0.02\pm0.02$  & $0.03\pm0.02$  & $0.04\pm0.03$  & $0.07\pm0.07$  & $0.07\pm0.07$\tabularnewline
\bottomrule
\end{tabular}
\par\end{centering}
\end{table*}

Note from Eq. \ref{eq:update_feed} that
if a neuron $n$ is mainly influenced by the external input at each
time $t$, then $x_{n}(t+1)\approx tanh(\epsilon\widetilde{u}_{n}(t))$ and
the corresponding states are distributed in a narrow
region around the hyperbolic tangent curve. This is what we see in Fig. \ref{fig:Activity-vs-filtered-1} for the non-plastic
case. Conversely, a broadened phase space (as found after plasticity
implementation) suggests a greater role of the interactions among past values of the reservoir units in determining the neuron state. In the case of neurons where IP was implemented, a further displacement of their activity towards the center of the activation
function is observed, which should come as no surprise since
we chose a zero-mean Gaussian as our IP target distribution. The
fact that mechanisms apparently so disparate exhibit similar effects
at the neuron and network level motivates the idea of synergistic learning involving both, synaptic and nonsynaptic plasticity, which has been extensively backed up also in biological systems \citep{hanse_associating_2008,mozzachiodi_more_2010}.

\begin{figure}
\begin{centering}
\includegraphics[width=\columnwidth]{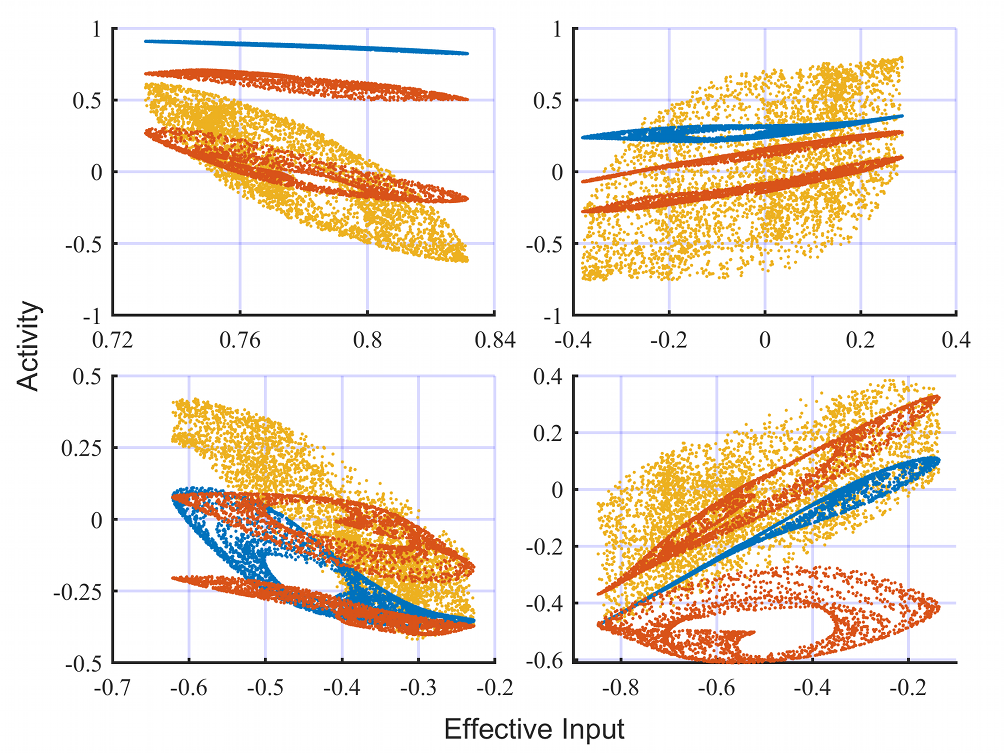} 
\par\end{centering}
\caption{Activity of 4 different neurons as a function of the effective input in a non-plastic ESN (blue) and in the same network after training it with NL anti-Hebbian rule for 25 epochs (red) and IP rule for 175 epochs (yellow).\label{fig:Activity-vs-filtered-1-1}}
\end{figure}

To finalize, we apply the same neuron-level framework to see if we
can understand the effects of an over-trained plasticity. It was shown in Fig. \ref{fig:Evolution-of-the} that once a certain number of epochs were exceeded, the prediction error increased, and that this co-occurred with a sharp increase in the spectral radius of the weight matrix for the NL anti-Hebbian rule (as shown in Fig. \ref{fig:Evolution-of-the-5}).
Is this observed transition from stable to unstable (periodic) dynamics reflected in any way in the activity of the neurons?
Choosing the same initial ESN
and training set as in Fig. \ref{fig:Activity-vs-filtered-1}, we now apply either the NL anti-Hebbian or IP rules for a total of $n_{Hebb}=25$, $n_{IP}=175$ epochs, respectively. From the resulting plot of the activity as a function of the effective input, as shown in Fig. \ref{fig:Activity-vs-filtered-1-1}, two different paths leading to the reported worsening of the prediction performance can be identified. On the one hand, the IP rule leads to a seemingly blurred phase space representation at each unit of the reservoir, in which each effective input value leads to a very spread network activity. We have identified that in this regime some neurons of the ESN lose their consistency (an important property that needs to be fulfilled in RC as discussed in \cite{bueno2017conditions,lymburn2019consistency}), producing different responses for the same input when the initial conditions are changed. On the other hand, an excess of NL anti-Hebbian training produces the split of the original phase space representation into two disjoint regions. We observed that the instability in this case is associated with a self-sustained periodic dynamics of the reservoir states, leading to consecutive jumps from one phase space region to the other. We have noticed that this transition, which results from the imposed decorrelation, is also followed by a decrease in the memory capacity of the network.

\section{Discussion and Outlook}

We showed that numerical implementation of plasticity rules
can increase both the prediction capabilities on temporal tasks and the memory capacity of reservoir computers. In the case of the Hebbian-type synaptic plasticity, we proved that a non-local
anti-Hebbian rule outperforms the classically used anti-Oja approximation. We also found that the synergistic action of synaptic (non-local anti-Hebbian) and non-synaptic (Intrinsic Plasticity) rules lead to the best results.

At the network level, different quantities that might be modified
by the plasticity rules were analyzed. For the non-local anti-Hebbian rule, we showed in Fig. \ref{fig:Evolution-of-the-5} how the sudden increase in the  reservoir weight matrix spectral radius co-occurs with a sharp drop on the states correlations at
consecutive times. More concretely, we observed that the optimal
number of epochs occurred just before the transition to a periodic self-sustained dynamics inside the reservoir. Similarly, continuous application of the IP rule also tends to decorrelate the states of the neurons within the reservoir. An over-training of the IP rule eventually results in a loss of the consistency of the neural responses and the corresponding performance degradation.

From the distributions of the states depicted in Fig. \ref{fig:Distribution-of-states},
we found that both types of plasticity lead to unimodal distributions
of states centered around zero. This seems to imply that the optimal
performance is achieved \textemdash for this type of temporal tasks\textemdash{}
when most neurons distribute along the hyperbolic tangent by avoiding the saturation regions.
Indeed, similar results observed both in vivo, on large monopolar cells of the fly \citep{laughlin_simple_1981}, and in artificial single neuron models \citep{bell_information-maximization_1995}, suggest that this form of states distribution helps to achieve optimal encoding of the inputs.

Interesting results emerged also from the one-to-one
comparison among individual neurons before and after the implementation of plasticity rules. At the neuron level, we saw how plasticity rules expand each neuron activity space \textemdash measured in terms of its area\textemdash, adapting to the properties of the input and thus possibly enhancing the computational
capability of the whole network. Within this same framework, we observed how the regime of performance degradation found when over-training the plastic parameters is of different nature for the synaptic and non-synaptic rules. In
the synaptic case, the phase space region occupied by the activity of each single neuron splits in two disjoint regions, with the state jumping from one region to the other at consecutive time steps. In the IP rule, on the other hand, we found that decorrelation of
the states and expansion of their phase space continues progressively, with different inputs eventually leading to similarly broad projections of the reservoir states in the activity phase space.

Our findings also raise interesting questions that will hopefully stimulate future works. On the one hand, we observe in Fig. \ref{fig:Activity-vs-filtered-1} that the resulting phase space after implementation of synaptic and non-synaptic plasticity is qualitatively similar for three out of the four neurons presented, while differing considerably from the non-plastic units. This remarkable result comes as fairly surprising
given that the synaptic and non-synaptic rules employed have indeed
very little in common from an algorithmic point of view. Nevertheless, they drive the network toward similar optimal states. On the other hand, the instability arising from an over-training of the plastic connections or intrinsic parameters shows to be of fundamentally different nature in the NL anti-Hebbian and IP rules. A thoughtful characterization of these transitions and a deeper understanding of the underlying similarities between synaptic and non-synaptic plasticity rules will likely trigger interesting research avenues.

The computational paradigm of reservoir computing has been shown to be compatible with the implementation constraints of hardware systems \cite{van2017advances,tanaka2019recent}.
The finding that a physical substrate with non-optimized conditions can be used for computation has been exploited in the context of electronic and photonic implementations of reservoir computing \cite{appeltant_information_2011,brunner2013parallel}.
Although the physical implementation of plasticity rules is certainly challenging, the results presented in this manuscript anticipate a potential advantage of considering such plasticity rules also in physical systems.

\section*{Declaration of Competing Interest}

The authors declare that they have no known competing financial interests or personal relationships that could have appeared to influence the work reported in this paper.

\section*{Acknowledgements} 

We thank M. A. Mu\~noz and M. A. Mat\'ias for fruitful scientific discussions.
This work was supported by MINECO (Spain), through project TEC2016-80063-C3 (AEI/FEDER, UE).
We acknowledge the Spanish State Research Agency, through the Severo Ochoa and Mar\'ia de Maeztu Program for Centers and Units of Excellence in R\&D (MDM-2017-0711).
The work of MCS has been supported by the Spanish Ministerio de Ciencia, Innovación y Universidades, la Agencia Estatal de Investigación, the European Social Fund and the University of the Balearic Islands through a "Ram\'on y Cajal” Fellowship (RYC-2015-18140). GBM has been partly supported by the Spanish National Research Council via a JAE Intro fellowship (JAEINT18\_EX\_0684).

\appendix

\section{Mackey-Glass series.\label{subsec:S.I.-1.-Mackey-Glass}}

The Mackey-Glass chaotic time series is generated from the following delay differential
equation: 
\begin{equation}
\dfrac{dx}{dt}=\left[\dfrac{\alpha x(t-\tau)}{1+x(t-\tau)^{\beta}}-\gamma x(t)\right],\label{eq:MG}
\end{equation}
where $\tau$ represents the delay and the parameters are set to
$\alpha=0.2$, $\beta=10$ and $\gamma=0.1$, a common choice for
this type of prediction tasks \cite{yusoff_modeling_2016,ortin2015unified}.

To construct the temporal series we solved Eq. \ref{eq:MG} using $Matlab$
dde23 delay differential equation solver, generating 10000 points
with an initial washout period of 1000 points. The step size between points
in the extracted series was set to 1, although it was initially computed
with step size of 0.1 and then sampled every 10 points. The absolute
error tolerance was set to $10^{-16}$, as suggested in \citep{jaeger_echo_2001}. All series were re-scaled to lay
in the range {[}0,1{]} before being fed to the ESNs. Similarly, the predicted points were re-scaled back to the original series range of values before computing the prediction accuracy.

We evaluated the prediction performance of the ESN for two values of the delay, 
$\tau=17$ for MG-17 and $\tau=30$ for MG-30, respectively.
For the MG-30 we generated $T=6000$ consecutive points
for the training set, while for the MG-17 we found that $T=4000$ training
points suffice to obtain good results. The evaluated prediction task consisted
on the continuation of the series from the last input of the training set. Accordingly, the target series in the supervised training was defined as the one-step-ahead prediction $y^{target}=[u_{2},u_{3},...,u_{T+1}]$
for an input $u=[u_{1},u_{2},...,u_{T}]$. For the computation of the output weights $W^{out}$, we kept all internal reservoir states of the ESN and only after passing all
the input training set we applied Eq. \ref{eq:Ridge}.

When implementing any of the plasticity rules, we ran the corresponding unsupervised learning procedure using the same $T$ points of the training mentioned above. The reservoir configuration was updated after every point of the input during this procedure. We passed over the whole training set a number of times (epochs).
 
At the end of the unsupervised learning procedure, the reservoir is kept fixed to the last configuration. We then collected the states
of the reservoir after a last presentation of the training set and
computed the new optimal readout weights using Eq. \ref{eq:Ridge}, as described in the previous paragraph.

\section{Derivation of anti-Oja rule.\label{subsec:S.I.-2.-Derivation}}

When applying a plasticity rule of the Hebbian type to ESNs, a common
choice is the anti-Oja rule \cite{yusoff_modeling_2016,babinec2007improving}. It is usually implemented using the following
local rule:
\[
w_{kj}(t+1)=w_{kj}(t)-\eta y_{k}(t)\left(x_{j}(t)-y_{k}(t)w_{kj}(t)\right).
\]where we denote the post-synaptic state of neuron $k$  by $y_k(t) \equiv x_k(t+1)$ in our RC scheme.

Originally, Oja proposed the plasticity rule that takes his name as
a way of deriving a local rule that implements Hebbian learning while
normalizing the synaptic weights at each step of the training. To
do so, he started from a common normalization of the basic Hebbian
rule:
\begin{equation}
w_{kj}(t+1)=\dfrac{w_{kj}(t)-\eta y_{k}(t)x_{j}(t)}{\sqrt{\sum_{j}\left(w_{kj}(t)-\eta y_{k}(t)x_{j}(t)\right)^{2}}},\label{eq:NL_Hebb}
\end{equation}
where we are already using the minus sign to account for anti-Hebbian
behavior. Notice that this form of the rule does not assume any particular
form of the activation function $y_{k}(t)=f(\vec{x})$. Now, this
update rule can be approximated by expanding the above expression
in powers of $\eta$:
\begin{flalign*}
w_{kj}&(t+1)=O(\eta^{2}) + \dfrac{w_{kj}(t)}{\sqrt{\sum_{j}\left(w_{kj}(t)\right)^{2}}}+ &\\ &+\eta\left[\dfrac{-y_{k}(t)x_{j}(t)}{\sqrt{\sum_{j}\left(w_{kj}(t)\right)^{2}}} 
+\dfrac{w_{kj}(t)2\sum_{j}\left(w_{kj}(t)\right)y_{k}(t)x_{j}(t)}{2\left(\sum_{j}\left(w_{kj}(t)\right)^{2}\right)^{3/2}}\right]&
\end{flalign*}

Imposing normalization of the incoming weights,
leads to
\[
w_{kj}(t+1)\sim w_{kj}(t)-\eta[y_{k}(t)x_{j}(t)-w_{kj}(t)y_{k}(t)\sum_{j}w_{kj}(t)x_{j}(t)].
\]
with ${\sqrt{\sum_{j}\left(w_{kj}(t)\right)^{2}}=1}$.

Finally, assuming linear activation functions and no external input,
so that $y_{k}(t)=\sum_{j}w_{kj}(t)x_{j}(t)$, we obtain the widely-known
anti-Oja's rule:
\begin{equation}
w_{kj}(t+1)\sim w_{kj}(t)-\eta\left(y_{k}(t)x_{j}(t)-w_{kj}(t)y_{k}(t)^{2}\right).
\label{eq:antiOja}
\end{equation}

The more adequate use of Eq. \ref{eq:NL_Hebb} comes of course with an
important computational cost compared to Eq. \ref{eq:antiOja},
but it is still feasible for the reservoir sizes we considered.

\section{Measures of reservoir dynamics during plasticity training.\label{subsec:S.I.-3.-Measures}}

To evaluate the decorrelation among pre and post-synaptic reservoir
states, we employed the Pearson correlation coefficient between activity
of unit $i$ at time $t$ and that of unit $k$ at time $t+1$, given
by: 
\begin{multline}
corr(x_{i}(t),x_{k}(t+1))=\\ \dfrac{\sum_{t=1}^{T-1}\left(x_{i}(t)-\overline{x}_{i}\right)\left(x_{k}(t+1)-\overline{x}_{k}\right)}{\sqrt{\sum_{t=1}^{T-1}\left(x_{i}(t)-\overline{x}_{i}\right)^{2}}\sqrt{\sum_{t=1}^{T-1}\left(x_{k}(t+1)-\overline{x}_{k}\right)^{2}}}
\end{multline}
After each epoch of the plasticity training, the mean absolute correlation
was computed as: 
\begin{equation}
\overline{Corr}=\dfrac{1}{M}\sum_{m=1}^{M}\dfrac{1}{N^{2}}\sum_{i=1}^{N}\sum_{k=1}^{N}\left|corr(x_{i}(t),x_{k}(t+1))\right|\label{eq:Correlation}
\end{equation}
where N denotes the size of the reservoir and M the number of independent
realizations over which the results were averaged.

\bibliographystyle{elsarticle-num}

\bibliography{mainrefs}

\vskip3pt

\bio{guillermo}
Guillermo B. Morales is currently working towards his PhD at the University of Granada. Previously, he received the MSC degree on complex systems at the University of the Balearic Islands. His main research interests cover topics of neural network dynamics and epidemics spreading from a complex systems perspective.
\endbio

\vskip3pt

\bio{claudio}
Claudio R. Mirasso is Full Professor at the Physics Department of the Universitat de les Illes Balears and member of the IFISC. He has co-authored over 160 publications included in the SCI with more than 7500 citations. He was coordinator (and principal investigator) of the OCCULT project (IST-2000-29683) and PHOCUS project (IST-2010-240763) and principal investigator of other national and European projects. His research interests include information processing in complex systems, synchronization, fundamentals and applications of machine learning, neuronal modelling and dynamics, and applications of nonlinear dynamics in general.
\endbio

\bio{miguel}
Miguel C. Soriano was born in Benicarlo, Spain, in 1979. He received the Ph.D. degree in applied sciences from the Vrije Universiteit Brussel, Brussels, Belgium, in 2006. He currently holds a tenure-track position at the University of the Balearic Islands, Spain. His main research interests cover topics of nonlinear dynamics and information processing based on reservoir computing. As an author or co-author, he has published over 60 research papers in international refereed journals.
\endbio

\end{document}